# High Velocity Penetration/Perforation Using Coupled Smooth Particle Hydrodynamics-Finite Element Method

S. Swaddiwudhipong[a,*], M.J. Islam[b], Z.S. Liu[c,*]
[a,b]Department of Civil Engineering, National University of Singapore, No. 1 Engineering Drive 2, Singapore 117576
[c]Institute of High Performance Computing, Fusionopolis Way, #16-16 Connexis, Singapore 138632

**ABSTRACT**

Finite element method (FEM) suffers from a serious mesh distortion problem when used for high velocity impact analyses. The smooth particle hydrodynamics (SPH) method is appropriate for this class of problems involving severe damages but at considerable computational cost. It is beneficial if the latter is adopted only in severely distorted regions and FEM further away. The coupled smooth particle hydrodynamics – finite element method (SFM) has been adopted in a commercial hydrocode LS-DYNA to study the perforation of Weldox 460E steel and AA5083-H116 aluminum plates with varying thicknesses and various projectile nose geometries including blunt, conical and ogival noses. Effects of the SPH domain size and particle density are studied considering the friction effect between the projectile and the target materials. The simulated residual velocities and the ballistic limit velocities from the SFM agree well with the published experimental data. The study shows that SFM is able to emulate the same failure mechanisms of the steel and aluminum plates as observed in various experimental investigations for initial impact velocity of 170 m/s and higher.

**Key words:** Element Distortion, Finite Element Method (FEM), High Velocity Impact, Perforation, Smooth Particle Hydrodynamics (SPH)

## 1. INTRODUCTION

The response of structures and materials subjected to dynamic loading (e.g. impact and blast loading of structures, vehicular impact, among others) has been a subject of interest for military, civil, automotive and aeronautical engineering. In particular, understanding of materials failure under high velocity ballistic impact is essential in the analysis and design of protective structures. Ballistic impact is a localized phenomenon where after impact projectile kinetic energy transforms into deformation and failure of both target and projectile. Ballistic limit velocity is the average of maximum and minimum projectile velocities for partial and complete penetrations respectively [1, 2], and it is one of the most common notions to indentify structures performance against projectile penetration. Backman and Goldsmith [1] showed several failure mechanism for thin and intermediate targets during ballistic penetration of projectile such as, spalling, petaling, plugging, ductile hole enlargement, fracture due to stress wave and fragmentation and failure in structures can cause by any of these failure mechanisms or through combination of more than one mechanisms.

*Corresponding authors. E-mail address: cvesomsa@nus.edu.sg; liuzs@ihpc.a-star.edu.sg

To date, significant improvement has been observed in the experimental studies for impact on metal targets at sub-ordnance (50–500 m/s) and ordnance (500–1300 m/s) velocity regimes [1–5]. Numerical studies are limited mainly due to the numerical instabilities observed in high velocity impact simulation. Nevertheless, a robust and efficient numerical approach is imperative for such problems since experimental studies usually require high cost and complex setups.

For high velocity projectile impact simulations, several numerical approaches such as finite element (FE) and mesh-free methods are available. The main challenge of the popularly used finite element simulations for high velocity penetration and/or perforation is the severe element distortion and damage observed in the target. These phenomena introduce numerical difficulties [6] leading to negative volume problem and premature termination of the analysis. The problem can be resolved by any of the following techniques: rezoning, element erosion, tunnel, local modified symmetry constraint or NABOR nodes algorithms [7]. The widely used approach is the element erosion method where severely distorted elements are removed from further analysis. The element erosion can be performed based on certain user defined failure criteria such as pressure, stress, strain [2, 8], damages and/or temperatures [4]. However, to the authors' knowledge, there are yet any direct approaches available to determine these erosion parameters.

Smooth particle hydrodynamics (SPH), a mesh-free method, is capable of handling large deformation without severe element distortion problem. Libersky et al. [9] adopted a 3D-SPH code MAGI to simulate the metal cylinder impact and hyper velocity impact tests. The results obtained were comparable to the experimental data. Since then the method has been adopted in a number of impact and fracture related problems. Liu et al. [10, 11] successfully employed the SPH method to study the dynamic response of structures under high velocity impact loads. The method is, however, usually less efficient computationally when compared to FEM and suffers from certain instability problems [12]. Combining both approaches, where the SPH is used to model the region of expected large deformation and damage, and FEM elsewhere [12], seems to be a logical development for high velocity projectile penetration and/or perforation simulation. In this study, 3D coupled SPH-FEM (SFM) has been used to simulate the perforation and/or penetration of steel and aluminum plates by steel projectiles. The study involves (i) the perforation of Weldox 460E steel plates by blunt projectile, with plate thickness varying from 6 to 20 mm, (ii) the perforation of 12 mm thick steel plates by projectiles with various nose shapes and (iii) the perforation of AA5083-H116 aluminum plates by conical steel projectile with plate thickness ranging from 15 to 30 mm. Simulated values of residual velocities and ballistic limit velocities are compared with experimental results.

## 2. MATERIAL MODEL

At high velocity impact, materials are subjected to extremely large strain, high strain rates, high temperature and severe damages. A constitutive relation for metals, Johnson-Cook (JC) model considering all the above parameters was proposed [13, 14]. The JC material model, damage parameter and fracture strain are expressed respectively as,

$$\sigma = \left[ A + B(\varepsilon_p)^a \right] \left[ 1 + C \ln \dot{\varepsilon}_{eff} \right] \left[ 1 - \left( (T - T_{room})/(T_{melt} - T_{room}) \right)^b \right] \quad (1)$$

$$D = \sum_{t=0}^{t_{cur}} \frac{\Delta \varepsilon_p}{\varepsilon^f} \quad (2)$$

$$\varepsilon^f = \left( D_1 + D_2 \exp D_3 \left( \sigma_{ave}/\sigma_e \right) \right) \left( 1 + D_4 \ln \dot{\varepsilon}_{eff} \right) \left( 1 + D_5 T^* \right) \quad (3)$$

In eqn. (1), $\varepsilon_p$ is the equivalent plastic strain, $\dot{\varepsilon}_{eff} = \dot{\varepsilon}_p/\dot{\varepsilon}_0$ is the effective strain rate, ($\dot{}$) implies differentiation with respect to time, $\dot{\varepsilon}_0$ (= 1 s$^{-1}$) and $\dot{\varepsilon}_p$ are the reference and plastic strain rates respectively, $T_{melt}$ and $T_{room}$ are the melting and room temperatures respectively, $t_{cur}$ is the time at the current step, $\sigma_{ave}$ and $\sigma_e$ are the average of normal stresses and Von Mises stress respectively, $A$, $B$, $a$, $C$ and $b$ are the material constants. The three brackets in eqn. (1) take into account the effects of plastic strains, the strain rates and the temperature respectively. Fracture in materials occurs by element erosion when $D$ is unity. $D_1$ to $D_5$ in eqn. (3) are the five damage parameters. This damage model is required as a remedial measure for severe element distortion problem in the FE simulation but it is not included for the SFM simulation in the present study.

## 3. NUMERICAL SIMULATION

Finite element method is widely used and well established. Both Lagrangian and Eulerian approaches are readily available commercially. Although the latter is more suitable for problems involving large deformation, it has certain disadvantages including difficulties in defining deformable material boundaries and the contact between the projectile and the target bodies, making the method inapt for ballistic penetration and/or perforation study [15]. Though Lagrangian formulation is easier to implement, it suffers from severe element distortion, notably in the vicinity under high velocity impact requiring a remedial measure.

Consider a three-dimensional (3D) body occupying a Lagragian space with volume $V$ subjected to traction $f_{t_i}(t)$ over a portion of outer surface $s_t$ and external body force $f_{b_i}(t)$. Virtual work principle requires that:

$$\int_V \rho \ddot{u}_i \delta u_i dV + \int_V \sigma_{ij} \delta u_{i,j} dV - \int_V \rho f_{b_i} \delta u_i dV - \int_{s_t} f_{t_i} \delta u_i ds = 0 \quad (4)$$

where $\rho$ is the material density, $\sigma_{ij}$ is the Cauchy's stress tensor, $\ddot{u}_i$ is the acceleration and $\delta u_i$ is the arbitrary virtual displacement. The comma implies the covariant differentiation. Applying finite element spatial discretization of eqn. (4), the governing equation becomes:

$$[M]\{\ddot{u}\} + [K]\{u\} = \{F\} \quad (5)$$

$$[M] = \sum_{n_0=1}^{n_{total}} \int_{V_e} \rho [N]^t [N] dV_e \quad (6)$$

$[M]$, $[K]$ and $[N]$ are the mass, stiffness and shape function matrices, $n_{total}$ is the total element numbers in the domain, $V_e$ is the element volume and $\{F\}$ is the equivalent nodal force vector of combined internal and external forces including those derived from the restitution of the bodies during the impact. For high velocity impact problem, the central difference explicit method in time is normally used to solve eqn. (5).

The smooth particle hydrodynamics method is a mesh-free Lagrangian method that can naturally handle problems involving large deformation and severe damaged materials and hence a suitable tool for high velocity impact studies. The method was first developed by Lucy [16], Gingold and Monaghan [17] to describe astrophysics phenomena. The system is represented by a set of particles, and the variables are calculated using the smoothing kernel functions. The integral representation or kernel approximation of a function $f(x)$ over a compact sub-domain of influence, $\Omega$, can be expressed as

$$\langle f(x) \rangle = \int_\Omega f(x_i) W(x - x_i, h) dx_i \qquad (7)$$

$W$ is the smoothing kernel function and $h$ is the smoothing length that is a unit measure of the sub-domain of influence of function $W$ (Fig. 1(a)). To satisfy the required partition of unity condition, the smoothing kernel function has to be normalized in each sub-domain:

$$\int_\Omega W(x - x_i, h) dx_i = 1 \qquad (8)$$

A commonly used smoothing kernel function is the cubic B-spline expressed as:

$$W(q, h) = \frac{\kappa}{h^\xi} \begin{cases} 1 - (3/2)q^2 + (3/4)q^3 & q \leq 1 \\ (1/4)(2-q)^3 & 1 < q \leq 2 \\ 0 & q > 2 \end{cases} \qquad (9)$$

where $q = (x - x_i)/h$, $\xi$ (= 1, 2 or 3) is the dimension of the problem and $\kappa$ is the scalar factor to comply with eqn. (8).

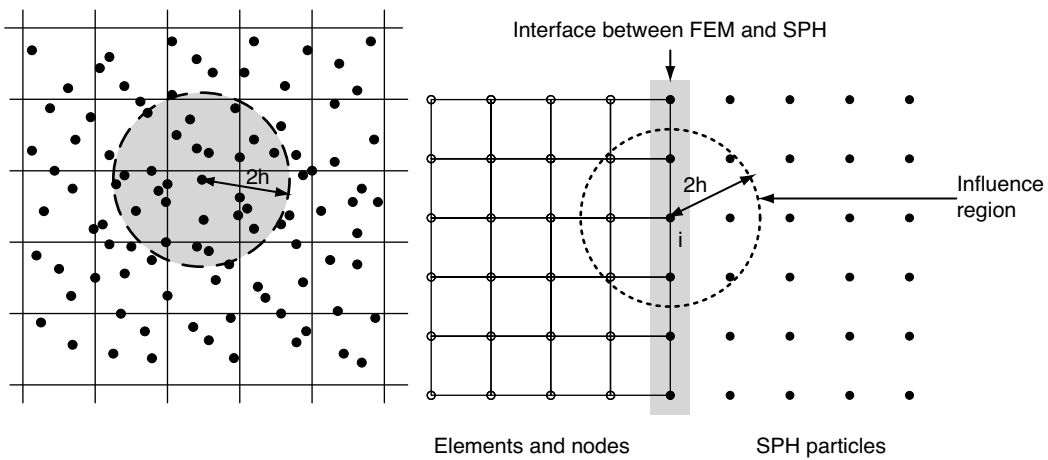

Figure 1. Coupling SPH and FE method (a) SPH particles with circular domain of influence and smoothing length ($h$), (b) Grid based elements and SPH particles

Based on the conservation of mass, momentum and energy in solid mechanics, the SPH governing equations are expressed as:

$$\frac{d\rho_i}{dt} = \sum_j \frac{\rho_i m_j}{\rho_j}(v_j^\beta - v_i^\beta)\frac{\partial W_{ij}}{\partial x^\beta} \tag{10}$$

$$\frac{dv_i^\alpha}{dt} = \sum_j m_j \left(\frac{\sigma_i^{\alpha\beta}}{\rho_i^2} + \frac{\sigma_j^{\alpha\beta}}{\rho_j^2}\right)\frac{\partial W_{ij}}{\partial x^\beta} \tag{11}$$

$$\frac{de_i}{dt} = \frac{\sigma_i^{\alpha\beta}}{\rho_i^2}\sum_j m_j(v_j^\alpha - v_i^\alpha)\frac{\partial W_{ij}}{\partial x^\beta} \tag{12}$$

where $m_j$, $\rho_j$, $\sigma_j^{\alpha\beta}$, and $v_j^\alpha$ are the mass, density, stress tensor and velocity associated with particle $j$, respectively, and $e_i$ is the energy per unit mass of particle $i$.

For high velocity impact, severe hydrostatic pressure is developed and usually evaluated via Mie-Gruneisen Equation of State (EOS) for solids. In the elastic regime, the deviatoric stress rate can be determined through Hooke's law, $\dot{S} = 2G\dot{\varepsilon}'$; but through the incremental plasticity theory for finite rotation using the Jaumann rate definition for post yield response. The Von Mises $J_2$ criterion and the associated flow rule are normally adopted to describe the plastic deformation in the type of target materials studied herein.

It is desirable to keep the number of particles and the mass in the influence sub-domain unchanged in both time and space for healthy numerical reasons. The total mass $M$ of a spherical influence sub-domain of radius $2h$ with $n$ number of particles can be expressed as:

$$M = n \times m = n \times \rho \times \frac{4}{3}\pi(2h)^3 = \frac{32}{3}n\rho h^3 \tag{13}$$

To keep the mass of the influence sub-domain unchanged, $dM/dt = 0$ and the conservation of mass requires that

$$\frac{d\rho}{dt} = -\rho\, div(v) \tag{14}$$

$$\frac{dh}{dt} = \frac{1}{3}h\, div(v) \tag{15}$$

Other inherent undesirable phenomena associated with standard SPH methods, such as tension instability and unstable execution caused by shock wave, have to be dealt with by a combination of remedial measures, e.g., introducing the additional stress method and the artificial viscosity term [10] to mitigate the above two deficiencies respectively. The field variables in the SPH governing equations, eqns. (10–12), can be directly updated using an explicit, leap-frog time integration algorithm.

To optimize the computational resources in the coupled SPH-FEM (SFM), the SPH particles are used in the region of expected large deformation and damages, while the rest of the domain is modeled by the finite element (FE) mesh. The SFM is able to reduce the computational resource requirements by reducing the number of SPH particles. Using the SPH method in only selected regions allows simulation of fractures and damages without any numerical problems. Moreover, using the FEM for the rest of the domain increases accuracy of the structural analysis.

Both SPH and FE methods are based on Lagrangian approach. Therefore, it is possible to link both methods at an interface as demonstrated in Fig. 1(b). The interface ensures continuous bonding of the two methods. At the interface, the SPH particles are constrained and moved with the elements. The influence sub-domain of the particles at/near the interface zone such as that of the particle "$i$" covers both the FE and SPH particles and certain considerations are required in the computation. For strain and strain rate calculation of each particle, only those from the SPH particles inside the influence sub-domain are considered, whereas the contributions from both SPH particles and elements inside the influence sub-domain are included to calculate the forces [12].

This study involves the SFM simulations of the perforation of Weldox 460E steel and AA5083-H116 aluminum plates of varying thicknesses impacted by projectiles of various nose shapes. Geometries of the three different nose shaped projectiles are depicted in Fig. 2 [4]. The modeling of each target plate comprises two regions. The SPH particles are adopted in the impact vicinity where damages and large deformation are expected while the rest of the target domain and the projectile are modeled using the FE 8-node solid elements with coarser mesh towards the outer boundary of the plate as illustrated in Fig. 3. Only a quarter of the problem is modeled using symmetry in $xz$ and $yz$ planes where symmetry boundary conditions are imposed for FEM mesh and a set of ghost particles are defined to enable the symmetry conditions for the SPH region. Numerical simulations are performed using hydrocode LS-DYNA [18]. The SPH particles and the finite elements surfaces are inter connected using a tied-nodes-to-surface contact feature. Contact between the projectile and the target plate is defined using an automatic-nodes-to-surface contact option.

Johnson-Cook (JC) material model is adopted for the target plates while each projectile is modeled as a simple elastic-plastic material with isotropic hardening. Though certain fragmentation and shattering of projectiles for thick steel plates were observed during the penetration, damages in the projectiles are not considered in the present study. The relevant material constants for steel and aluminum target plates and hardened steel projectile are listed in Tables 1–3 respectively.

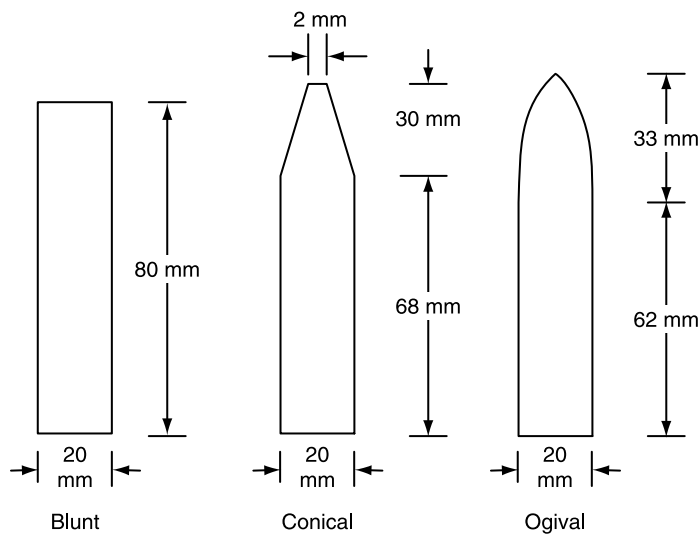

Figure 2. Geometry and dimension of the various nose shaped projectiles

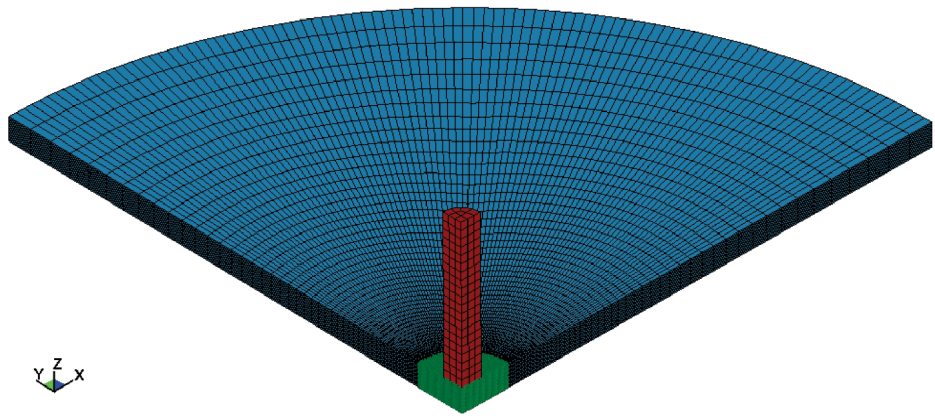

Figure 3. Mesh of the target and projectile numerical model

Table 1. JC Material properties for Weldox 460 E steel plate (4)

| $\rho_0$ (kg/m³) | $E$ (GPa) | $v$ | $G$ (GPa) | $A$ (MPa) | $B$ (MPa) |
|---|---|---|---|---|---|
| 7850 | 210 | 0.33 | 75 | 499 | 382 |
| $a$ | $C$ | $b$ | $C_p$(J/kgK) | $T_{melt}$(K) | $T_{room}$(K) |
| 0.458 | 0.0079 | 0.893 | 452 | 1800 | 293 |
| $D_1$ | $D_2$ | $D_3$ | $D_4$ | $D_5$ | |
| 0.636 | 1.936 | −2.969 | −0.014 | 1.014 | |

Table 2. JC Material properties for AA5083-H116 aluminum plate (5)

| $\rho_0$ (kg/m³) | $E$ (GPa) | $v$ | $G$ (GPa) | $A$ (MPa) | $B$ (MPa) |
|---|---|---|---|---|---|
| 2700 | 70 | 0.3 | 27 | 167 | 596 |
| $a$ | $C$ | $b$ | $C_p$(J/kgK) | $T_{melt}$(K) | $T_{room}$(K) |
| 0.551 | 0.001 | 0.859 | 910 | 893 | 293 |

Table 3. Material properties for hardened Arne tool-steel (4)

| $\sigma_Y$(GPa) | $\rho_0$(kg/m³) | $E$ (GPa) | $v$ | $E_t$(GPa) | Mean $\varepsilon_f$ (%) |
|---|---|---|---|---|---|
| 1.9 | 7850 | 204 | .33 | 15 | 2.15 |

### 3.1. DOMAIN SIZE SENSITIVITY STUDY
The choice of SPH domain size is studied to ensure the adequacy of the SPH region and the economy of computational resources. Deformation region governs the SPH domain size. Zukas [19] mentioned a severe deformation zone of 3–6 times the projectile diameter for ballistic impact cases. Three SPH domain radii of 24, 30 and 36 mm are adopted in this study expecting a severe deformation zone of 2.4–3.6 times the projectile diameter. Domain size sensitivity studies are performed on blunt projectile perforation of two Weldox 460E steel plates of thicknesses 8 and 16 mm. Numerical residual velocities of the projectiles are compared with the experimental results as shown in Fig. 4. As all three sets of results show good convergence for both cases, the SPH domain radius (*r*) of 24 mm is used in subsequent analyses. Normally, twice the projectile diameter is adequate for the SPH domain size for the SFM.

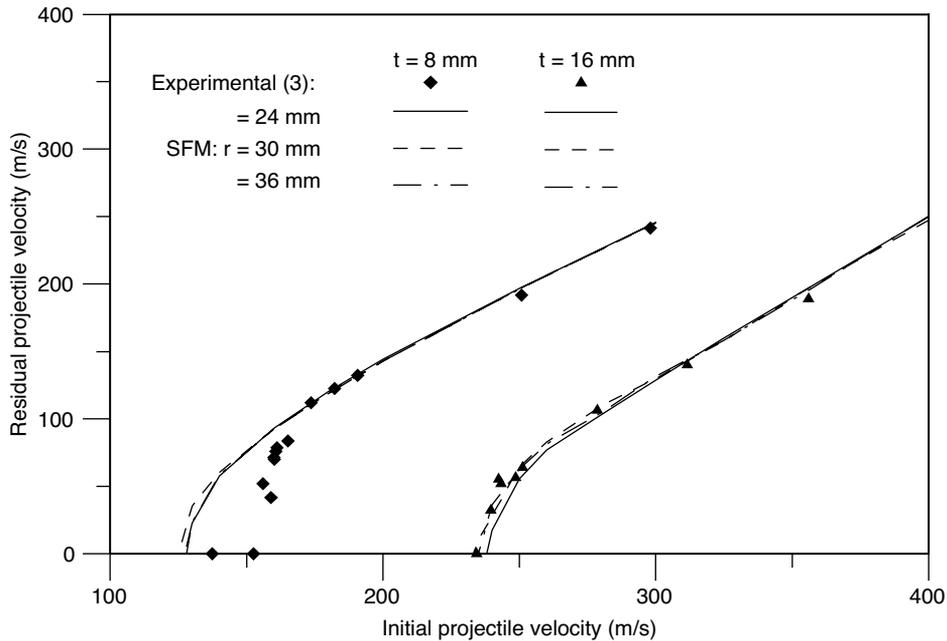

Figure 4. Domain size sensitivity study for steel plates perforated by blunt projectile

### 3.2. EFFECT OF SPH PARTICLE DISTANCE
Initial numerical results of blunt projectile perforation of steel plate using SFM showed that the residual velocities are sensitive to the SPH particle distance. The phenomenon of mesh sensitivity is also observed for FE simulation by Dey [4] who stipulated that it is due to the localized adiabatic shear failure around the periphery of the projectile. Therefore, the study is conducted to study the effects of the SPH particle distance for two sampled plate thicknesses of 8 and 16 mm. The results from the SPH convergence study of the two cases as shown in Fig. 5 illustrate that reasonable convergent results can be achieved using the SPH particle distance of 0.6 mm and the value is adopted in subsequent computations. The effects of the SPH particle distance are also studied for sharp nose projectiles and the results indicate the same particle distance of 0.6 mm to be adopted.

### 3.3. EFFECT OF FRICTION
The melting temperature and the strength of the target material affect the values of the friction coefficient to be used in the study. The lower melting temperature tends to induce a thin layer between the target and projectile that acts as a lubricant. The photomicrograph of the penetration of aluminum target plates by a spherical nose projectile at 1120 m/s initial velocity showed significant microstructural changes in a thin layer of 5–15 $\mu$m in the target around the projectile [20]. Similar behavior was also observed for other sharp such as conical and ogival nose projectiles. At the contact surface between the target and the projectile, the target materials flow both up and down. The phenomenon suggests that sliding friction between the projectile and the target exists and has to be considered. No such layer was observed for blunt projectile perforation as the target plate failed by localized adiabatic shear failure [4] inducing negligible contact friction between the projectile and the target. This is confirmed by the observed constant residual velocity after the failure of the target by adiabatic shear and plugging of blunt projectile as reported earlier by Børvik et al. [3].

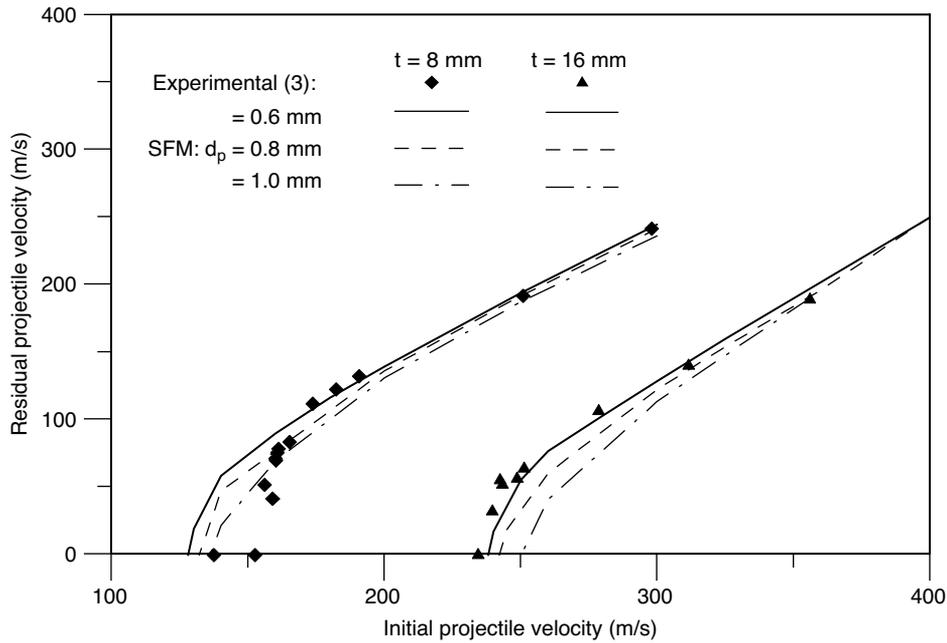

Figure 5. SPH particle distances ($d_p$) sensitivity study for steel plate perforation by blunt projectile

Selecting a proper value of the friction coefficient, $\mu$, is not easy as no direct experimental data are readily available for high velocity impact. Ravid and Bodner [21] assumed the values of $\mu = 0.1$ and $\mu = 0.05$ in high velocity rigid projectile perforation of steel plates for frontal and lateral projectile surfaces respectively. Lower value for the lateral projectile surface was expected due to the effect of high velocity and the presence of thin viscous film as material temperature rises beyond the melting point at contact surfaces. Three different values of friction coefficients of 0.05, 0.08 and 0.1 are used to study the perforation of the conical nose steel projectile into the steel target plate. The residual versus the initial projectile velocity plots adopting the above three $\mu$ values as shown in Fig. 6 illustrate a significant effect of friction on the residual velocity. The in-between value of 0.08 for $\mu$ seems to provide reasonably accurate results simulated via SFM and is adopted in subsequent simulations of conical and ogival nose projectile perforations.

Forrestal et al. [20] suggested the values of friction coefficients of 0.02 to 0.20 for sharp nose steel projectile penetration into 6061-T651 aluminum targets. Fig. 6 compares the experimental data with those obtained based on the $\mu$ values of 0.0, 0.02 and 0.05 for 15 mm thick AA5083-H116 aluminum plate perforation by conical nose steel projectile. The numerical results using the value of 0.02 as suggested by Montgomery [22] are observed to agree well with those from the impact tests. This value is adopted for subsequent aluminum plate perforation studies.

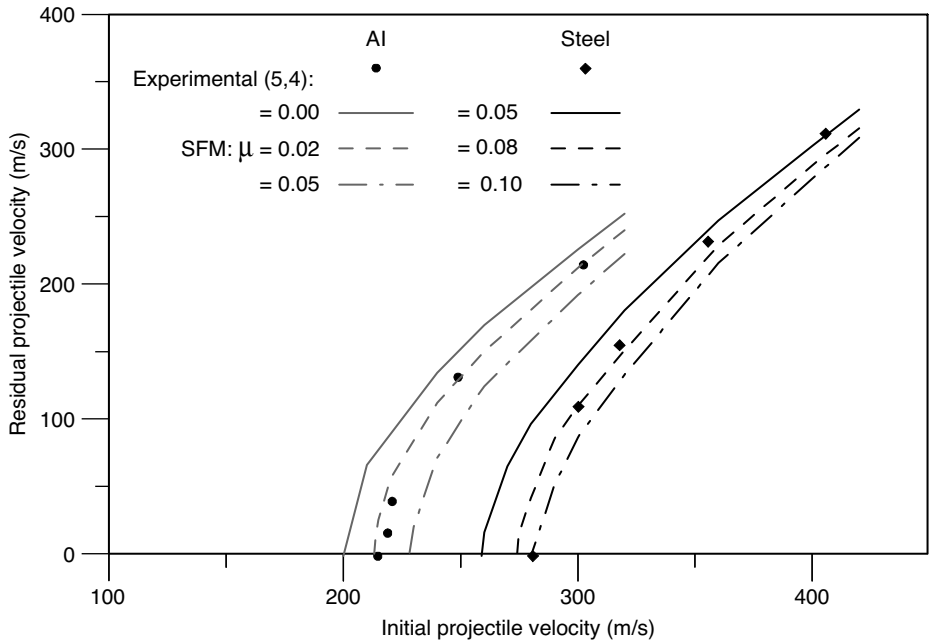

Figure 6. Effect of friction in conical projectile perforation of Al and steel plates

## 4. PERFORATION OF WELDOX 460E STEEL PLATES
### 4.1. BLUNT PROJECTILE PERFORATION

In the present study, perforations of steel plates with thickness ranging from 6 to 20 mm by blunt projectile using 3D SFM in LS-DYNA have been carried out. Numerical residual and ballistic limit velocities are compared with the experimental data from Børvik et al. [3]. Fig. 7 shows the residual velocity versus the initial velocity plots for various plate thicknesses. Except for thin plates at relatively low initial projectile velocities of about 170 m/s and less the SFM results agree well with experimental values. During the experiment, Børvik et al. [3] observed a sudden drop in the projectile residual velocities for perforation of 6 and 8 mm thickness target plates. However, this is not apparent in numerical solutions and the residual velocities for various plate thicknesses are well distributed.

Ballistic limit velocity is defined as the minimum projectile velocity needed to penetrate the whole target plate. The SFM provides a good representation of the experimental results for plate thickness of 10 mm and above but seems to underestimate the experimental ballistic limit velocities for those of 8 mm and below. Ballistic limit velocities for various plate thicknesses of 6 mm to 20 mm are illustrated in Fig. 8. A certain change in the slope of the curve is evident for the experimental results at a plate thickness of 10 mm. This difference in slope was explained by Børvik et al. [3] as the change in failure mode from adiabatic shear and plugging failure for thick plates to global dishing and plugging failure for thin plates.

Similar study is performed using 3D finite element method (FEM). To avoid the severe element distortion problem in FEM, damage based element erosion method is used for the target plates. Based on the convergence study, the element of $0.25 \times 0.25 \times 0.25$ mm$^3$ is adopted to model the target plate in the impact vicinity and the mesh is gradually coarser towards the outer edge. The FEM results, along with the experimental and SFM values as displayed in Fig. 8, predicts a change in slope similar to those observed in the experiments.

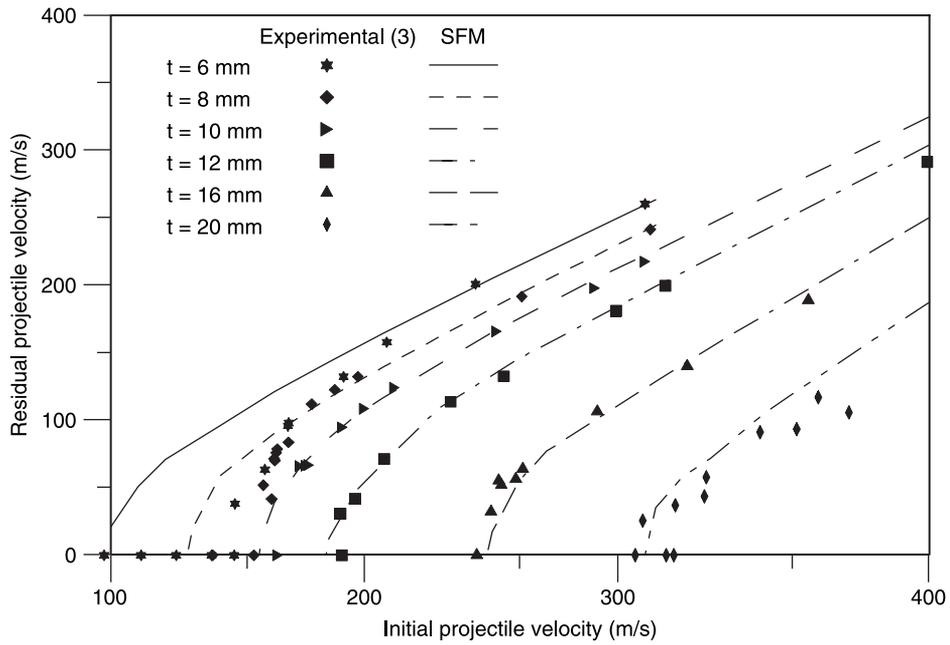

Figure 7. Numerical and experimental residual velocities for blunt projectile perforating steel plates

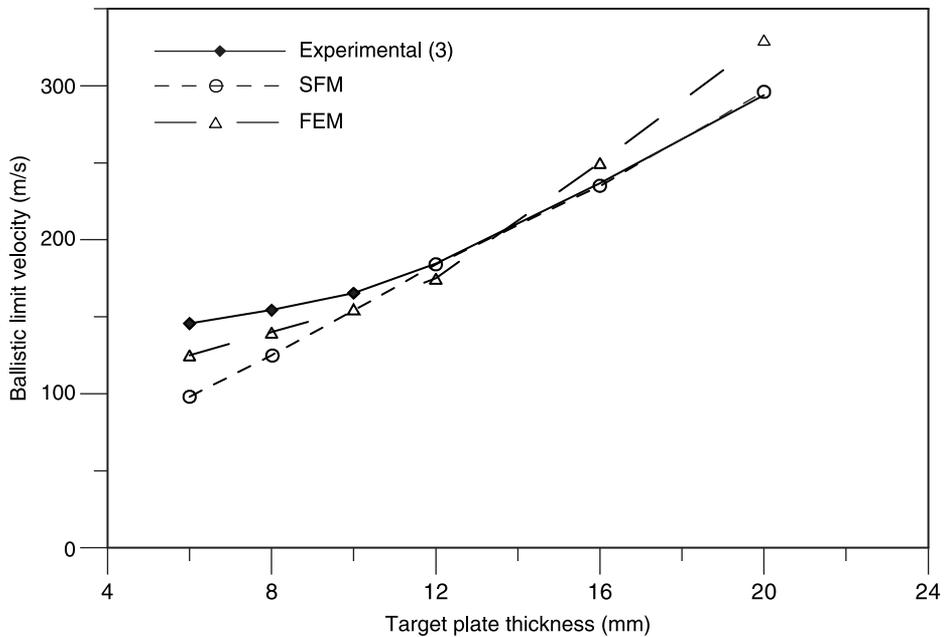

Figure 8. Numerical and experimental ballistic limit velocities for blunt projectile perforating steel plates

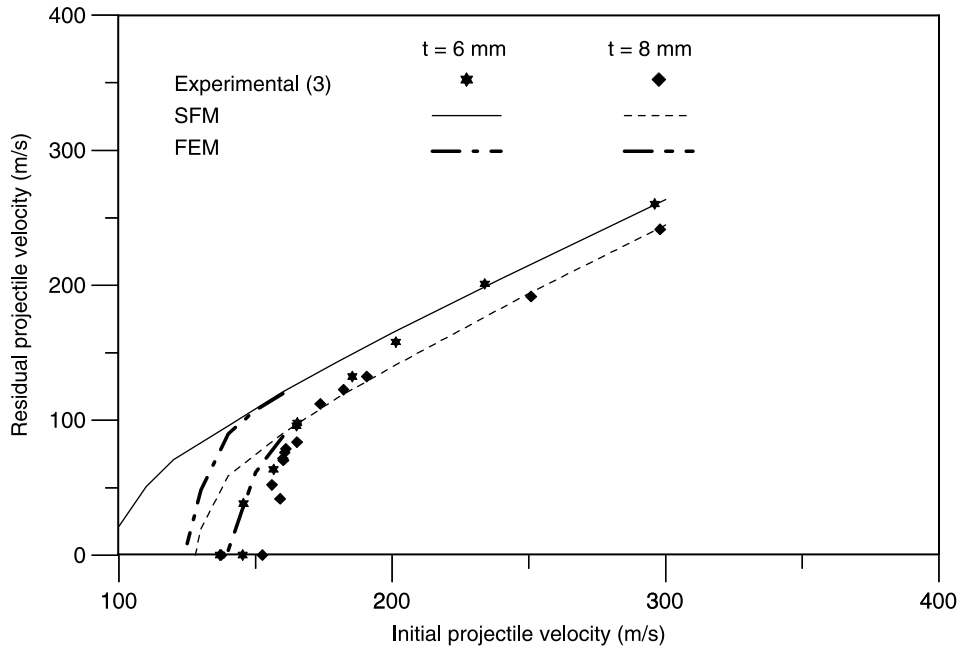

Figure 9. Better performance of FEM as compared to SFM at low initial velocity perforation

The SFM residual velocity versus the initial velocity plots for 6 and 8 mm thick plates are shown in Fig. 9. The FEM results for initial velocity of 170 m/s and less are also included for comparison. The perforation by blunt projectiles at lower initial projectile velocity (170 m/s or less) seems to be better simulated via FEM, while the SFM simulation performs better at higher initial velocities. Because of the severe element distortion and target damages in FE model, the FEM requires four to five times more computational time than those of the SFM for these cases. Appropriate numerical approach for relatively thin plates, depending on the level of impact velocities, should be judiciously selected.

Numerical study is carried out for 12 mm thick plate perforation using only smooth particle hydrodynamics (SPH) method in the target domain. Projectile is modeled as usual with FEM. Because of the large target domain size, the system restricts the use of more particles at 0.6 mm particle distance for the SPH part. There remain two options, either use a smaller target domain size or adopt a larger particle distance. Since smaller domain size may cause boundary effects, the latter option is selected for the current study. A particle distance of 2.0 mm is chosen for the SPH target part. Comparison of the SPH and SFM results is shown in Fig. 10. Although the SPH solutions differ from the experimental and SFM results, they show the same trend. A smaller particle distance would provide a closer agreement. However, computational resource requirements for the SPH method analysis are substantially higher than those of the SFM.

### 4.2. PERFORATION BY PROJECTILES OF VARIOUS NOSE GEOMETRIES

The perforation of 12 mm thick Weldox 460E steel plates by blunt, conical and ogival nose shaped projectiles are carried out. Numerical residual and ballistic limit velocities are compared with the experimental data from Dey [4]. The numerical results agree well with the experimental data as shown in Fig. 11. Table 4 shows that the SFM ballistic limit velocities for these cases deviate less than 6% from the experimental values. Failure patterns of the target plates due to perforation of the three different projectile nose geometries as illustrated in Fig. 12 are similar to the experimental observations reported by Dey [4]. For blunt projectile, the failure of the plate is via adiabatic shear and plugging modes with a plug thickness close to that of the plate. The spherical and conical projectiles are observed to progress through each target by moving material in the radial direction and ductile hole enlargement with petal pattern detected at the rear surface.

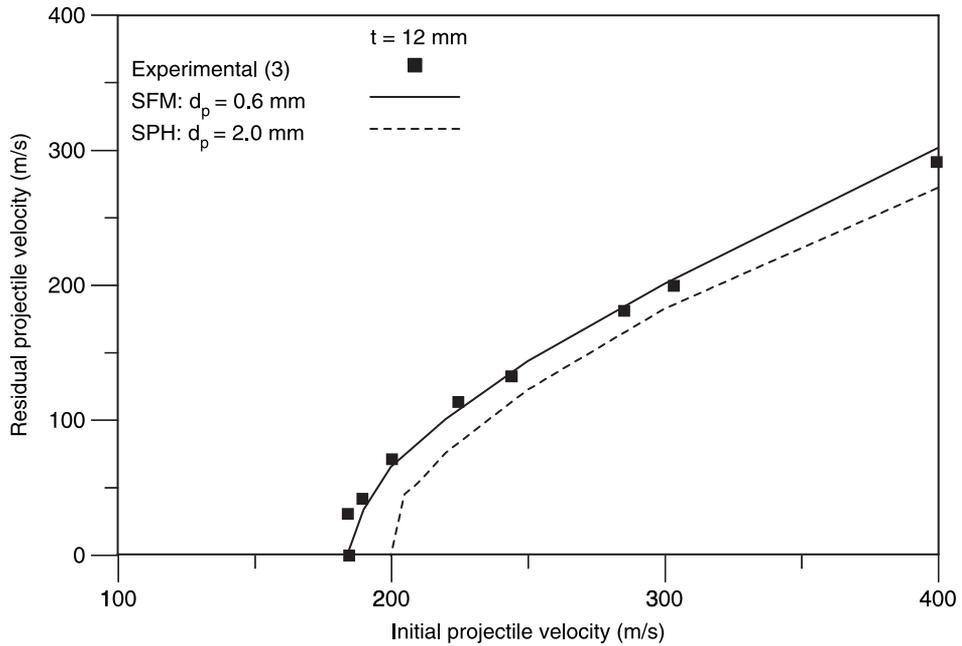

Figure 10. Performance study of SFM and SPH method for steel plate perforation by blunt projectile

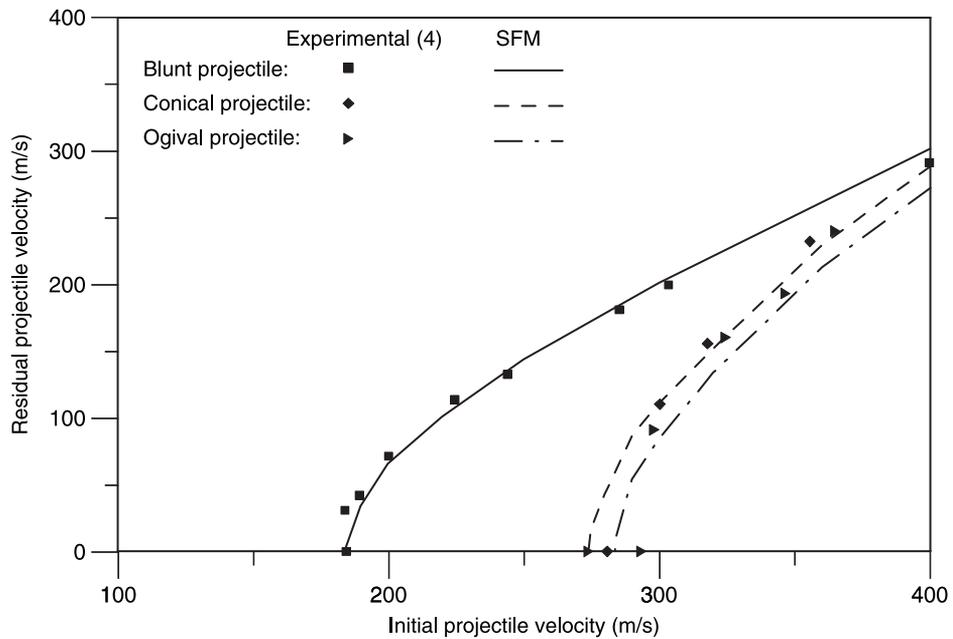

Figure 11. Comparison of numerical and experimental residual velocities

Table 4. Ballistic limit velocity ($V_{bl}$) for three different projectiles

| | Ballistic limit velocity, $V_{bl}$ (m/s) | | |
| --- | --- | --- | --- |
| | **Blunt** | **Conical** | **Ogival** |
| Experimental [4] | 184.5 | 290.6 | 295.9 |
| CSPHFEM | 184.0 | 275.0 | 284.0 |

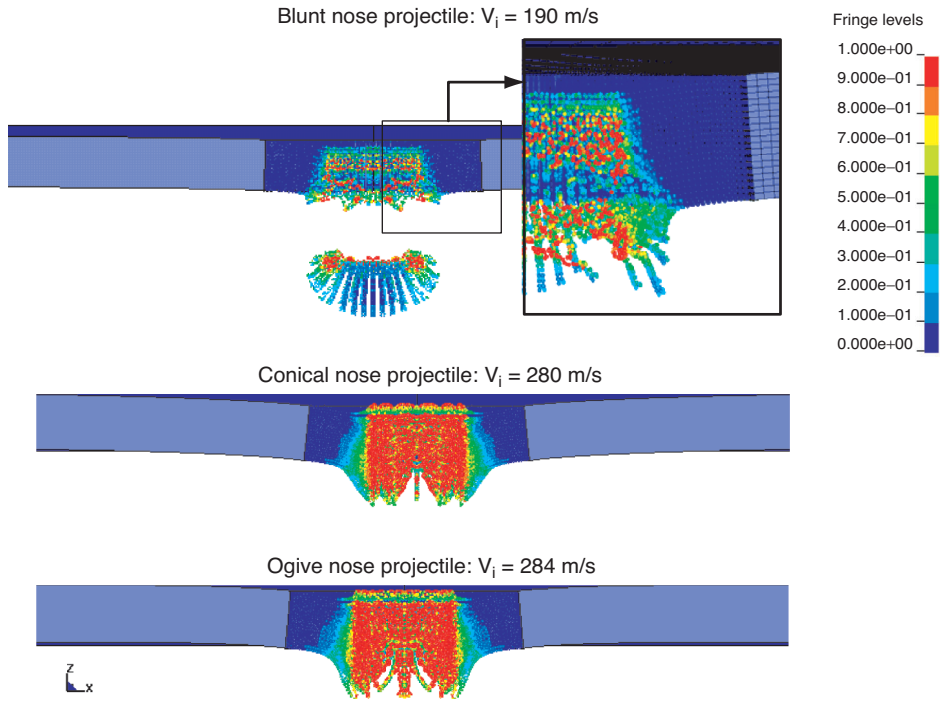

Figure 12. Steel plates after perforation showing effective plastic strain fringe contour

## 5. PERFORATION OF AA5083-H116 ALUMINUM PLATES

Perforation of AA5083-H116 aluminum plates with thickness ranging from 15 mm to 30 mm by conical nose projectile using 3D SFM in LS-DYNA have been performed. Numerical residual projectile velocities and ballistic limit velocities are compared with the experimental data reported earlier by Børvik et al. [5]. Variation of the residual velocities with the initial velocities for different plate thicknesses of 15, 20, 25 and 30 mm are presented in Fig. 13. Ballistic limit velocities increase linearly with increasing plate thicknesses as shown in Fig. 14 indicating a similar failure pattern for all plate thicknesses. The SFM results show a good agreement with those observed in the experiments. Fig. 15 presents the discrete time history of the perforation process of 15 mm thick aluminum plate perforated by a conical projectile. The plate fails due to ductile hole enlargement, forming petals at the rear surface of the target plates. Similar behavior is observed for all other plate thicknesses. Fig. 16 illustrates the target plates after perforation of the projectiles at or near the ballistic limit velocities. Petals are observed and the failure patterns are consistent with the experimental observation [5]. The fringe contour of the effective plastic strain confirming the confinement of plastic deformation within the SPH portion is also included in Fig. 16.

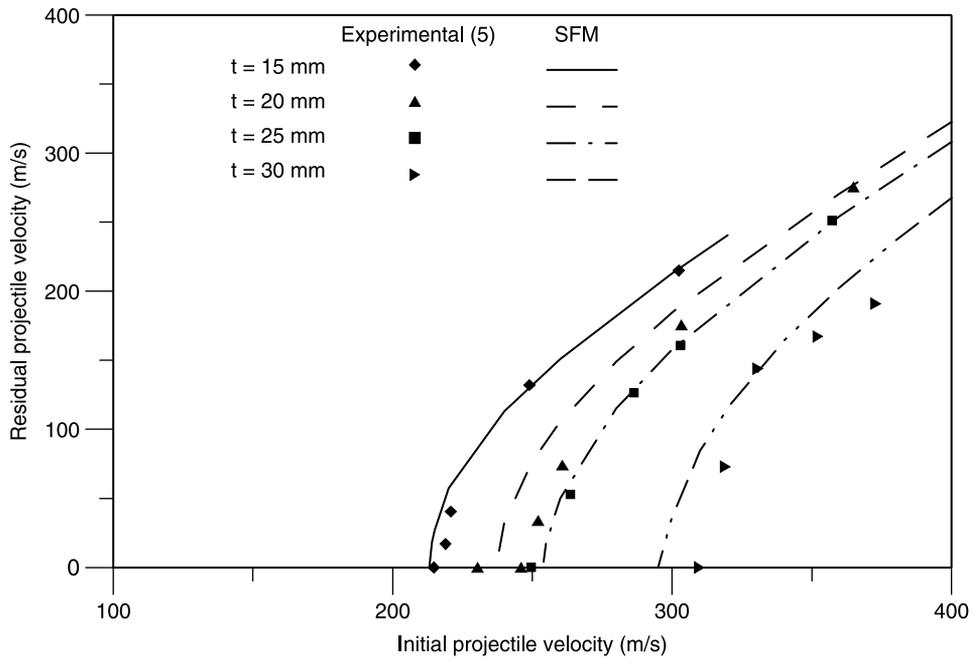

Figure 13. Numerical and experimental residual velocities of conical projectiles perforating aluminum plates

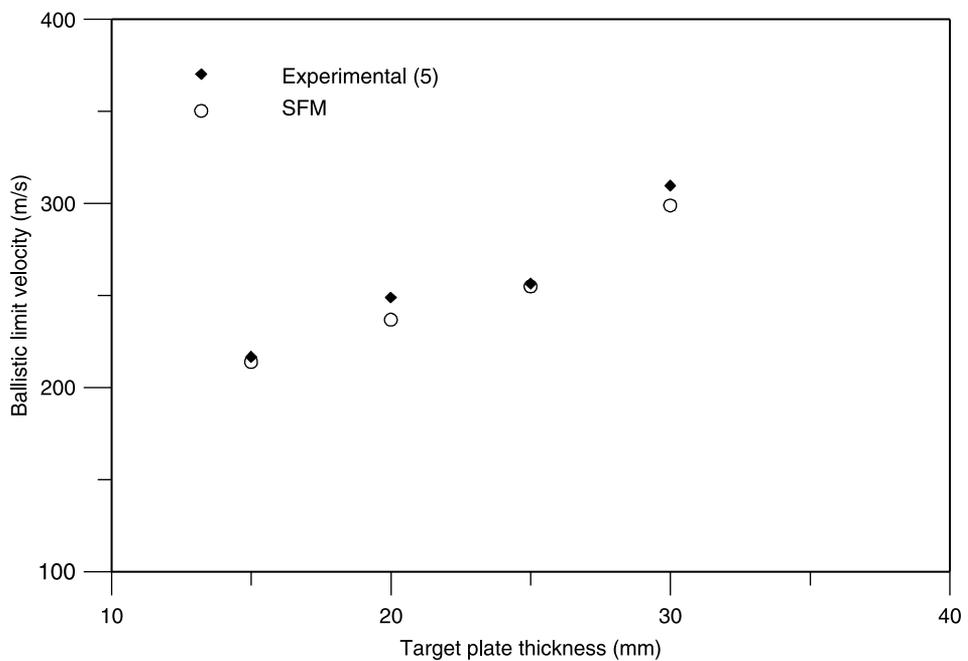

Figure 14. Numerical and experimental ballistic limit velocities for aluminum plate perforation

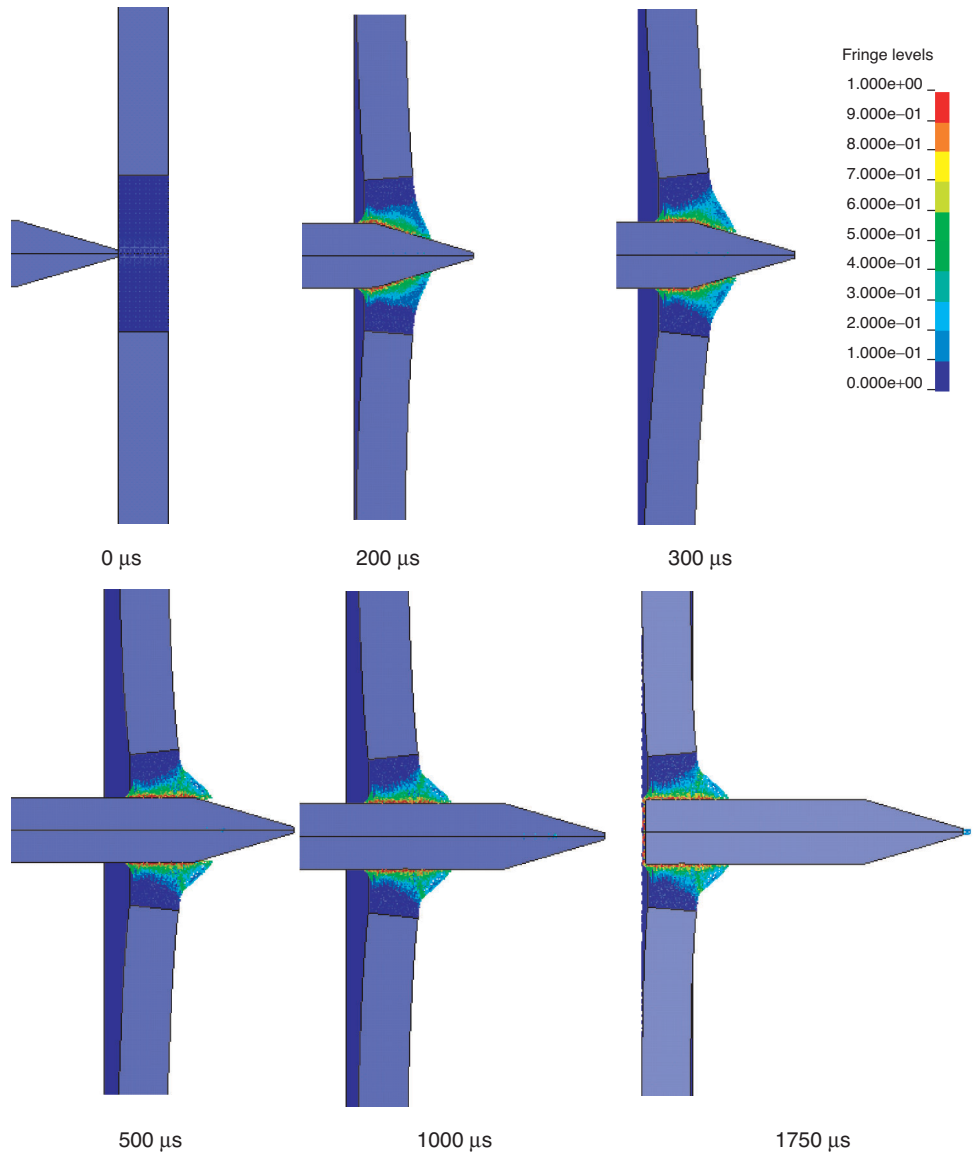

Figure 15. Time history of 15 mm thick aluminum plate perforated by conical projectiles at $v_i$ = 214 m/s with effective plastic strain fringe contour

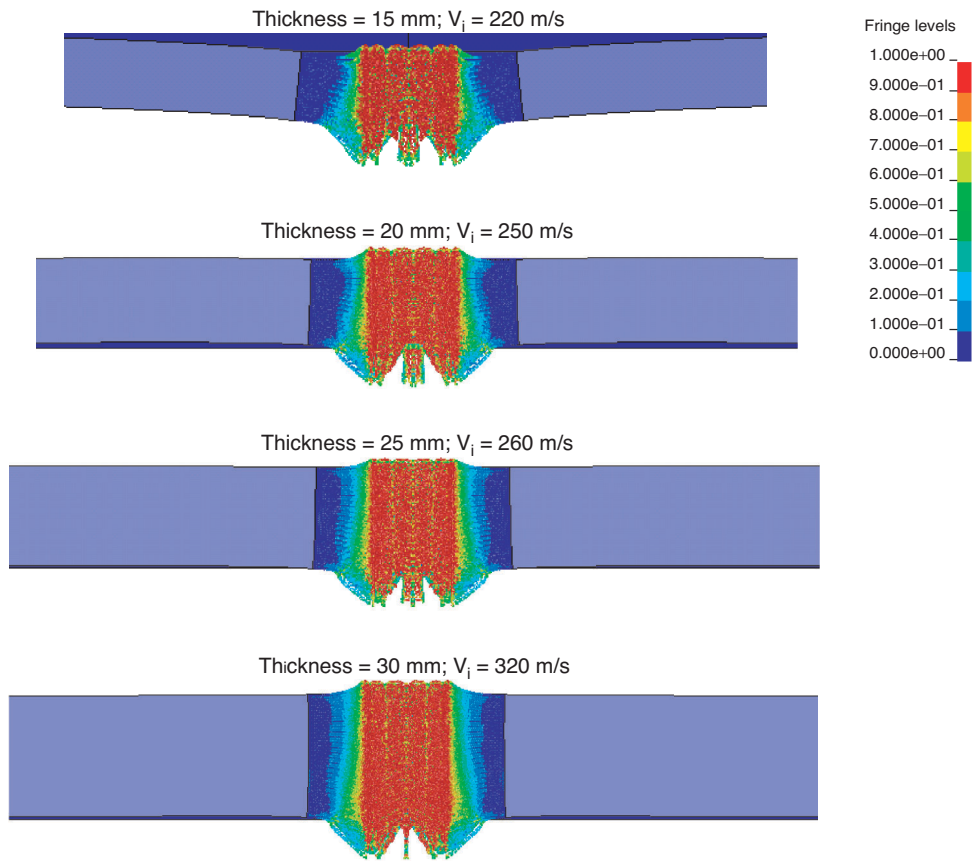

Figure 16. Aluminum plates after perforation by projectile at/near ballistic limit velocities with effective plastic strain fringe contour

## 6. CONCLUSIONS

The coupled SPH-FEM (SFM) is adopted to simulate high velocity perforation of steel and aluminum plates of different thicknesses perforated by steel projectiles with various nose geometries. The method is able to predict rather accurately the modes of failure, the projectile residual velocities and ballistic limit velocities as compared with those observed in the test reported earlier except for those due to blunt projectile impact at low velocity of 170 m/s or less. This deviation in results is observed for the perforation of thin plates as the change in failure pattern is not reflected in the solution obtained from the adopted method at low impact velocity on thin plates due mainly to the tensile instability problem inherent in the SPH method. At lower range of impact velocities, FE solutions are in better agreement and may be adopted for this range of impact velocities. The SFM combines the strength of SPH and FEM methods while addresses their short falls of computational demanding and early program termination due to severe element distortion respectively. Though the SFM is less accurate at low velocity impact of 170 m/s and lower, the method is robust and efficient for high velocity impact penetration and/or perforation of both steel and aluminum target plates.